\documentstyle[a4,12pt,amsfonts,amssymb,amscd]{report}

\begin{document}
\begin{titlepage}

\title{On Polyakov's basic variational formula for loop spaces}
\author{Ioannis P. Zois\thanks{izois\,@\,maths.ox.ac.uk}\\
\\
Mathematical Institute,\\ 24-29 St. Giles', Oxford OX1 3LB}
\date{}
\maketitle

\begin{abstract}
 We use the homological algebra context to give a more 
rigorous proof of Polyakov's basic variational formula for loop
spaces.\\
PACS classification: 11.10.-z, 11.15.-q, 11.30.Ly
\end{abstract}
\end{titlepage}

\section{Introduction-Motivation}

 It is known for some time now that one can reformulate Yang-Mills
theory as \emph{non-linear $\sigma $ model} (abriviated to
\emph{"nl$\sigma $m"} in the sequel) on the \emph{loop space}
\cite{polyakov}. This was related originally to the problem of \emph{confinement
of quarks} \cite{polyakov}, \cite{wilson}. In addition recently a
fascinating "electric-magnetic" duality was observed in the loop space
formulation of Yang-Mills theories \cite{tsou}. 

 Apart from the obvious disadvantages one has when formulating a field
theory (in particular nl$\sigma $m here) on an infinite dimensional
space (namely the loop space of the initial manifold), there are
however some simplifications in field equations and an extra $U(1)$
symmetry (coming from rotating loops) and hence this
approach is not merely an extra undesired nuance \cite{polyakov},
\cite{tsou}. There are also some mathematical advantages related to
the Duistermat-Heckman formula \cite{atiyah}, \cite{witten} and to
the heat equation proof of the Atiyah-Singer index theorem \cite{bismut}. This reformulation is based on Polyakov's basic
variational formula \cite{polyakov}:
$$\delta h(c)=\int _0^1 ds P\left(exp\int _0^s A_{\mu}dx^{\mu}\right)F_{\mu \nu
}(c(s))\frac{dx^{\nu }(s)}{ds}P\left(exp\int _s^1 A_{\mu}dx^{\mu}\right)\delta
x^{\mu}(s)$$
where $F $ is the curvature of a connection 1-form $A$ on spacetime,
$c$ is a loop, $h(c)$ is the holonomy element:
$$h(c)=Pexp\int _{c} A$$
and $P$ is the well-known \emph{Dyson ordering}. The loop $c$ is
described via the function $x^{\mu}(s)$.

 Using the isomorphism induced by the \emph{iterated integral map}
between the \emph{Hochschild homology} of the associative algebra of
differential forms of the
original manifold and the \emph{de Rham cohomology} of the
corresponding loop space when the manifold is simply connected, (see
\cite{GJP}, \cite{jones}, \cite{chen}), we give a more mathematically rigorous proof
of Polyakov's result.

 We hope moreover that some ideas and techniques from iterated
integrals will be of some use to the canonical quantization of gravity
using Astekhar variables because loop spaces are also important in
that context.

 In more concrete terms, let $M$ be a four dimensional simply connected manifold (assumed to be spacetime) and let $LM$
be its loop space, namely the set of smooth maps from the circle $T$
to $M$. The dimension of $M$ is not crucial, it can be anything, we
choose four due to physical significance. What is crucial is that $M$
has to be \emph{simply connected}. In quantum field theory context
usualy spacetime has the topology of ${\bf R}^4$ with either Euclidian
or Minkowski metric. We denote by $\Omega (M)$ the associative algebra
of differential forms on $M$. Then the above mentioned result simply
states that if $M$ is simply connected then the Hochschild homology of
$\Omega (M)$ is isomorphic to the de Rham cohomology of $LM$. The
isomorphism is the one induced by the iterated integral map. As a
reference for Hochschild homology, see \cite{loday}.

 We organise this paper as follows: In section 2 we briefly present Polyakov's main
ideas; in section 3 we give a brief
review of loop spaces. Since iterated integrals are not, we think, \textsl{lingua franca} in
physics, we give in section 4 the basic definitions. After that we give the
proof itself in section 5. Then we end up with some remarks in section
6.

\section{Yang-Mills theory as a nl$\sigma $m on Loop space}

 (The main reference in this section is \cite{polyakov}).

 In establishing the above formulation, the basic role is played by a
well-known object, the element of the holonomy group.  Following standard physics terminology we write the holonomy element $h$
as
$$h(c)=Pexp\int _{c} A$$
where $c$ is some loop, $A$ is a connection 1-form and $P$ stands for the \emph{Dyson ordering} along this
loop. We now consider $h$ as a chiral field. In mathematical
language $h$ is a zero form on the loop space
of $M$ with values in the group $G$. The underlying mathematical
structure is a principal bundle $X$ over $M$ with structure Lie group
$G$ assumed to be compact and connected. We introduce a connection
$\tilde{A}$ on the loop
space by the formula
$$\tilde{A}_\mu (s,c):=\frac{\delta h}{\delta x^{\mu } (s)} h^{-1}$$
where 
$$\frac{\delta }{\delta x^{\mu }(s)}$$
is the loop derivative \cite{tsou}.

In the above expression the loop $c$ is parametrised by the function
$x^{\mu}(s)$ and clearly the index $\mu $ takes the values
0,...,3. The above defined connection should be reparametrisation
invariant because $h(c)$ is and hence one must have

$$\frac{dx^{\mu}(s)}{ds}\tilde{A}_{\mu}(s,c)=0$$

 From the definition of the connection on the loop space one can
deduce that this connection is flat \cite{polyakov}, namely

$$\frac{\delta \tilde{A}_{\mu}(s,c)}{\delta x^{\nu}(s_1)}-\frac{\delta
\tilde{A}_{\nu}(s_1,c)}{\delta x^{\mu}(s)}+[\tilde{A}_{\mu}(s,c),\tilde{A}_{\nu}(s_1,c)]=0$$

We would like to note here that strictly speaking this is not a
connection on the G-bundle over the loop space since it is defined
explicitly in terms of the given holonomy and there is no notion of
"gauge transformation" on $\tilde{A}_{\mu }(s,c)$ itself. However for
convenience we shall refer to it as a connection.

 The important result is that the Yang-Mills equations take  a
simple form in terms of the connection on the loop space \cite{polyakov}, namely

$$\frac{\delta \tilde{A}_{\mu}(s,c)}{\delta x^{\mu}(s)}=0$$

So to sum up, one has two important results when formulating
Yang-Mills equations on loop spaces:

1. The connection 1-form on the loop space is \emph{flat} even if the
space-time connection one starts with is not.\\

2. The Yang-Mills equations reduce to a "divergenceless-like" condition
for the connection 1-form on loop space. (Actually the word "like" is
very important, one cannot define a Hodge star on the loop space forms
since the de Rham complex is not bounded above---due to infinite dimensionality---although one has a
naturally induced metric on the loop space if the manifold itself has
a metric, see below).

 These two results mentioned above are based on the
following variational formula:

$$\delta h(c)=\int _0^1 ds P\left(exp\int _0^s A_{\mu}dx^{\mu}\right)F_{\mu \nu
}(c(s))\frac{dx^{\nu }(s)}{ds}P\left(exp\int _s^1 A_{\mu}dx^{\mu}\right)\delta
x^{\mu}(s)$$
where $F$ is the curvature of the connection $A$ on space time. Our
proof explains the appearence of the curvature in this formula quite naturally.\\

\section{Loop spaces}

In this section we review some well-known facts about loop spaces in
general. More details can be found in \cite{atiyah}.

 Consider a finite dimensional compact orientable Riemannian manifold $M$. Then by
definition the loop space of $M$ is the following infinite dimensional
manifold

$$LM:= Map (T,M)$$
consisting of all smooth maps from the circle $T$ to our manifold. Our
description of $LM$ will not be absolutelly rigorous, we ignore
analytic issues.

 Thus a point on $LM$ is by definition a smooth map $\phi
:T\rightarrow M$ and the tangent space $T_{\phi}LM$ of LM at $\phi $
can be identified with the space of sections of the vector bundle
$\phi ^*(TM)$, the tangent bundle $TM$ of $M$ pulled-back to $T$ by
$\phi $. The metric on $M$ defines a metric on $\phi ^*(TM)$ and hence
by integration over $T$ we get an inner product on the  space of
sections. This defines a pre-Hilbert space structure on
$T_{\phi}LM$. Next we introduce the Levi-Civita connection $\nabla $
on $M$. This induces a connection on the bundle $\phi ^*(TM)$ and hence
(evaluating along the tangent vector to the loop) a covariant
derivative operator $\nabla _{\phi}$. This is a skew-adjoint operator
on the space of sections $T_{\phi}LM$ and hence, using the inner
product, it defines a skew-bilinear form on $T_{\phi }LM$. As we now
vary the point $\phi \in LM$ we get a 2-form on $LM$. One can prove that
it is closed, the proof based crucially on the use of the Levi-Civita
connection on $M$. However this form is not non-degenerate in general, this
is so only at the points $\phi $ for which $\nabla _{\phi }$ has a zero
eigenvalue, i.e. a tangent vector to $M$ which is covariantly constant
along the loop $\phi $. The Hamiltonian function $H$ associated to the
obvious action of the circle can nonetheless still be defined as
follows: recall that the energy $E$ of a loop $\phi $ is defined as 

$$E(\phi )=\frac{1}{2}\int _{T} |d\phi |$$

 Computing the derivative of $E$ in the direction of a tangent vector
$\xi \in T_\phi $ we get

$$(dE,\xi )=\int _{T}<\frac{d\phi }{dt},\nabla _{\phi }\xi >$$
which establishes that $E=H$. This allows one to get an analogue of
the Duistermaat-Heckman formula in infinite dimensions. 

The lesson here therefore is that the loop
space of any Riemannian manifold is \emph{almost} a symplectic manifold and
in fact most of the "symplectic" things can be done on the loop space
(if one can overcome the infinite dimensions!).

 The orientability of $LM$ can be understood as follows: we have the
natural evaluation map $f:T\times LM\rightarrow M$; pulling back by
$f^*$ and then integrating over $T$ induces a homomorphism

$$a:H^2(M;{\bf Z}_{2})\rightarrow H^1(LM;{\bf Z}_{2})$$
 The image of the second Stiefel-Whitney class of $M$ is then the
obstruction to orientability of $LM$. In particular, if $M$ is spin, then
$LM$ is orientable. The converse holds if $M$ is also simply connected. 
 
 We would like to mention another fact which is not relevant for our immediate
discusion but it is useful to know and it is actually one of the main
motivations to study loop spaces in general: the Wiener integration on the loop
space $LM$ is related to the heat equation on $M$, thus giving, 
another way to calculate the index of elliptic operators on $M$ using
data from $LM$. One however must be careful to distinguish between the
Wiener measure (using Riemannian structure) and the Liouville meassure
(using symplectic structure) in the case of $LM$. They are related via
the Pfaffian. This fact is also useful in physics, in SUSY nl$\sigma
$m.(cf \cite{witten}).

\section{Iterated integrals}

 We start with some motivation first. Consider the following ODE

$$\frac{d\phi (t)}{dt}=a(t)\phi (t)$$
where $a(t)$ is a given function and we want to solve this in the
interval [0,1] given the initial condition $\phi (0)=1$. This is
trivial. Yet one may ask the following non trivial question: is it possible to calculate the single \emph{value} $\phi (1)$ \emph{without}
solving the equation with respect to the function $\phi (t)$?

 The answer is \emph{yes} and the formula is the following:

$$\phi (1)=\sum _{k=0}^{\infty }\int _{\Delta _k}
a(t_1)...a(t_k)dt_1...dt_k$$
where $\Delta _k$ is the standard k-simplex 
$$\{(t_1,...,t_k)\in {\bf R}^k :
0\leq t_1\leq ...\leq t_k\leq 1\}$$
The above is a sum of iterated integrals, namely each term $k$ \emph{is} an
iterated integral.

One can easily notice that the above equation suitably generalised,
actually gives the correct expression for parallel transport of a
vector field (replacing $\phi $ above) given a connection 1-form $A$
(replacing $a(t)$ above), namely essentially the covariant
derivative. And now we think the whole thing starts to take shape. In the
above setting then, assuming [0,1] parametrising a circle, the value
$\phi (1)$ is exactly the holonomy (more precisely the final vector
which is the holonomy times the initial vector as matrices).
This
is actually the observation that made us think about relating
Polyakov's formulae on loop spaces  with iterated integrals.
 Let us however, before giving formal definitions, write down an
iterated integral explicitly: suppose we are in Euclidian space
$\bf{R}^n$. If $w=w_i(x)dx^i$ and $v=v_i(x)dx^i$ are two real valued
1-forms on $\bf{R} ^n$ and suppose $a:[0,1]\rightarrow \bf{R} ^n$ is a path, then
the "twice" iterated integral of the forms $w$ and $v$ is by definition
the following expression:

$$\int _0^1[\int _0^t w_i(x(a(t)))dx^i(a(t))]v_j(x(a(t)))dx^j(a(t))$$

We now pass directly to the definitions on the loop spaces. We shall
begin with some general facts about spaces carrying smooth circle
actions. (We write them specifically for the loop spaces, but they hold
in general for spaces carrying smooth circle actions).

 The loop space $LM$ may
be given the structure of an infinite dimensional manifold  modeled on
a Frechet space. The circle group acts smoothly by rotating loops,
namely $(\phi _{t}c)(s)=c(s+t)$ where $c(s)$ is a
loop and $\phi _{t}$ is a smooth 1-parameter group of diffeomorphisms
with period 1 which describes the smooth circle action. This natural circle action on $LM$ defines several
operators on the space of differential forms on $LM$. The first is the
contraction with the vector field which is tangent to the loop, namely
the generator of the T-action, which will be denoted by $i$. Then
there is an averaging operator $\Theta $ defined via

$$\Theta (w)=\int _0^1 \phi _{t}^{*}wdt$$
 Furthermore there is a sequence of operators $p_k$ defined as follows:

 $$p_{k}:\Omega
(LM)^{\otimes k}\rightarrow \Omega (LM)$$
where $1\leq k<\infty $
 
 The explicit formula is the following: given a form $w$ on $LM$, let $w(t)$
be the form $\phi _t^*(w)$ on $LM$; then one has:

$$p_k(w_1,...,w_k)=\int _{\Delta _k}iw_1(t_1)\wedge ...\wedge iw_k(t_k)dt_1...dt_k$$
where $\Delta _k$ is the standard k-simplex 

$$\{(t_1,...,t_k)\in {\bf R}^k | 0\leq t_1\leq ...\leq t_k\leq 1\}$$

 We shall now give the key property of the maps $p_k$ :\\

\textbf{Proposition:}\\

\textit{ If $\epsilon _{i}=|w_1|+...+|w_k|-i$, then}\\

$$dp_{k}(w_1,...,w_k)=-\sum _{i=1}^k (-1)^{\epsilon
_{i}-1}p_k(w_1,...,dw_i,...w_k)$$
$$-w_{1}p_{k-1}(w_2,...,w_k)$$
$$-\sum _{i=1}^{k-1}(-1)^{\epsilon
_i}p_{k-1}(w_1,...,w_{i}w_{i+1},...,w_k)$$
$$+(-1)^{\epsilon _{k-1}}p_{k-1}(w_1,...,w_{k-1})w_k$$

 This formula simply states the fact that 

$$\sum _{k=0}^{\infty }p_k$$
is a Hochschild cocycle on the differential graded algebra (DGA for short) $\Omega (LM)$ with coefficients in
$\Omega (LM)$ itself.\\

 The proof of the above Proposition is by direct computation using the explicit formula we gave
for the maps $p_k$ above (cf \cite{GJP}).\\

 One can observe that $p_1$ in particular is equal to $i\Theta =\Theta
i$ and that its square is
zero. Moreover it anticommutes with the de Rham differential $d$. Hence
one actually has a \emph{mixed complex} $(\Omega (LM),d,p_1)$. This
observation will be important later. 

 We now pass to the iterated integrals. Notice first the shifting of the degree between forms on
$M$ and forms on $LM$ by the following example: if $w$ is a 1-form on $M$,
then $\int _{c}w$ is a function on $LM$, where $c\in LM$. Iterated integrals
generalise this idea. If $w$ is a form on $M$, let $w(t)$ be the form
$e_t^*(w)$ on $LM$, namely the pull back of $w$ via the evaluation map
$e_{t}:LM\rightarrow M$ given by evaluating loops at time $t$. Given
forms $w_0, w_1, ..., w_k$ on $M$, the iterated integral 

$$\sigma (w_0,...,w_k)$$
is a form on $LM$ of total degree $|w_0|+...+|w_k|-k$ defined by the
formula 

$$\sigma (w_o,...,w_k)=\int _{\Delta _k}w_0(0)\wedge iw_1(t_1)\wedge ...\wedge
iw_k(t_k)dt_1...dt_k$$
where $\Delta _k$ is the standard k-simplex and $i$ is the contraction
operator with the tangent vector to the loop.
 
We can rewrite the above formula for the iterated integral using the
maps $p_k$, namely  

$$\sigma (w_0,...,w_k)=w_{0}p_k(w_1(0),...,w_k(0))$$

One can then build a model for the forms on $LM$ using iterated
integrals. We begin by recalling the definition of the \textbf{cyclic
bar complex} \cite{adams} of the algebra $\Omega (M)$. Let $C(\Omega (M))$ be the
direct sum

$$\sum _{k=0}^{\infty} \Omega (M)\otimes s\Omega (M)^{\otimes k}$$

Here $s$ is the suspension functor on graded vector spaces, that is the
functor which simply reduces degree by 1. In general, the cyclic bar
complex of any associative algebra comes naturally equipped with two "differentials", the
\emph{Hochschild} differential $b_0$ defined via
\\
$$b_0(w_0,...,w_k)=-\sum _{i=0}^{k-1}(-1)^{\epsilon
_i}(w_0,...,w_{i-1},w_{i}w_{i+1},w_{i+2},...,w_k)$$
$$+(-1)^{(|w_k|-1)\epsilon
_{k-1}}(w_{k}w_0,w_1,...,w_{k-1})$$
and  \emph{Connes'} differential $B$ defined via

$$B(w_0,...,w_k)=\sum _{i=0}^{k}(-1)^{(\epsilon _{i-1}+1)(\epsilon
_{k}-\epsilon _{i-1})}(1,w_i,...,w_k,w_0,...,w_{i-1})$$
$$-\sum _{i=0}^{k}(-1)^{(\epsilon _{i-1}+1)(\epsilon _{k}-\epsilon
_{i-1})}(w_i,...,w_k,w_0,...,w_{i-1},1)$$

 However, in our case we are interested in the cyclic bar complex of
the algebra $\Omega (M)$ which is itself also a DGA with de Rham
differential $d$. In the bar complex then one has an extension of this
$d$, still denoted $d$, given via

$$d(w_0,...,w_k)=-\sum _{i=0}^{k}(-1)^{\epsilon
_{i-1}}(w_0,...,w_{i-1},dw_i,w_{i+1},...,w_k)$$  
As before, $\epsilon _{i}=|w_0|+...+|w_i|-i$\\

 Now we combine the de Rham differential $d$ on $C(\Omega (M))$ with the
Hochschild differential $b_0$ on $C(\Omega (M))$ to get a single
differential $b$:

$$b=d+b_0$$

We shall refer to this cohomology $(C(\Omega (M)),b)$  as the \emph{Hochschild
cohomology} of $\Omega (M)$ Now one also has a \emph{mixed complex} for the cyclic bar complex
$C(\Omega (M))$, namely $(C(\Omega (M)),b,B)$. The total differential
$b+B$ gives the \emph{cyclic cohomology} of $\Omega (M)$
 
 And now we are ready to state the main results relating iterated
integrals, forms on loop space $LM$ and the cyclic bar complex of the
algebra of forms on $M$ (see \cite{GJP} for proofs and further explanations on
notation and terminology):\\

\textbf{Theorems :}

\textit{1. The iterated integral map $\sigma $ induces a map between the two
mixed complexes}  

$$(C(\Omega (M)),b,B)\rightarrow (\Omega (LM),d,p_1)$$

This simply means that $p_{1}\sigma =\sigma B$ and
$\sigma b=d\sigma $. (For the proof see \cite{GJP}).\\

\textit{2. If $M$ is simply connected, one has an isomorphism between the de
Rham cohomology of the loop space and the Hochschild cohomology of the
 algebra of forms on $M$ induced by the
iterated integral map $\sigma $, namely}

$$\sigma :(C(\Omega (M)),b)\rightarrow (\Omega (LM),d)$$
\textit{is an isomorphism in cohomology}. (Again for the proof see \cite{GJP}
or \cite{chen}).\\

\textit{3. "Essentially" the cyclic cohomology of $\Omega (M)$ is isomorphic to the
$T$-equivariant cohomology of $LM$. (The word "essentially" means that we
ignore the complications that lead to the correct Jones' variant $HC^{-}_{-*}$
functor)}. (For the proof see \cite{jones}).\\

\textit{4. Under the map $\sigma $, the shuffle product on the normalised
cyclic bar complex $N(\Omega (M))$ of $\Omega (M)$ is carried into the wedge product
on $\Omega (LM)$}. (For the proof and terminology see \cite{GJP}).\\

\textit{5. The forms on $LM$ which are images of the iterated integral map are
basic with respect to the action of the Lie pair
($vect[0,1],Diff[0,1]$). In particular this means that they are
\emph{reparametrisation invariant} under reparametrisations of the
loops} \cite{GJP}.\\

 We just want to end this section by mentioning that many of the above
generalise to actions of arbitrary compact Lie groups $G$ acting on
manifolds. One then gets \emph{equivariant} versions of the above
results. This is actually what we need in our physical problem,
because we consider gauge theories and \emph{Lie algebra valued} forms
whereas all the above discussion refers to \emph{real} valued
forms. Fortunately, if one starts with a principal bundle $P$ over a
base manifold $M$ with structure group $G$, the loop
space $LP$ also is a principal G-bundle over $LM$ in a natural way, see
\cite{bismut}. This actually implies that the generalisations are
straightforward.

\section{The Proof}

 To begin with, in physics people usually work with \emph{based}
loops, which means that one picks a point $x\in M$ and considers loops
having this point as starting and ending point. We shall denote this
space $LM(x)$. The reason for this is that one can compose loops easily
in this way. 

 This is not actually important in our treatment because we shall use
formulae valid for the bigger space of free loops. We must however
mention that here we consider smooth loops whereas in physics one can
consider more general loops which are only continuous.

 First one writes the holonomy element $h$ as an infinite sum of
iterated integrals by expanding the Dyson ordering. We use the formula of the $\sigma $ map in terms
of the maps $p_k$ where we conventionally define $p_0=1$ and we assume
the 0th form $w_0(0)$ appearing in the formula to be equal to the constant
form 1 and thus we omit
it, since there is no integration on that either. 

 In more concrete terms then
$$h=\sum _{n=o}^{\infty }p_n(A_1,...,A_n)$$
where we simplify the notation slightly by writing $A_i$ instead of
$A_i(0)$, $1\leq i\leq n$. In our notation the index $i$ states the
\textsl{"position"} of the form $A$. Recall that because this is the holonomy element, namely an
element of the structure group $G$, we know that the above converges.

 Anyway, then one goes on by looking at Polyakov's variational formula and it
is not hard to suspect that this "looks like" taking the $d$ of the
holonomy element $h$. The $d$ enters the sum and hits every individual
term $p_n$. The
interchange of $d$ and $\sum $ is justified because both sides make
sense. In fact by definition the $d$ of the $p_n$ is
actually a sum of $p_n$'s, applied to different forms (see formula
 of Proposition mentioned above). Moreover recall that as mentioned in \cite{GJP}, the sum $\sum
_{n=0}^{\infty }p_n$ is a Hochschild cocycle no matter what forms it
is applied to and also recall that each iterated integral is finite, hence the sum
is well defined (converges):

$$dh=d\sum _{n=0}^{\infty }p_n(A_1,...,A_n)=$$

$$=\sum _{n=o}^{\infty }dp_n(A_1,...,A_n)=$$

We now take each term separately and applying the formula of
Proposition  above we get:

$$dp_0=0$$\\
$$dp_1(A_1)=p_1(dA_1)$$\\
$$dp_2(A_1,A_2)=p_2(dA_1,A_2)+p_2(A_1,dA_2)-p_1(A_1\wedge A_2)$$\\
$$dp_3(A_1,A_2,A_3)=p_3(dA_1,A_2,A_3)+p_3(A_1,dA_2,A_3)$$
$$+p_3(A_1,A_2,dA_3)-p_2(A_1\wedge
A_2,A_3)-p_2(A_1,A_2\wedge A_3)$$\\
$$+...$$
 
 Introducing the curvature 2-form $F$
of the connection 1-form $A$ to be $F=dA-A\wedge A$, we have the
folowing formula:

$$dh=p_1(F)+$$
$$+p_2(F_1,A_2)+p_2(A_1,F_2)+$$
$$+p_3(F_1,A_2,A_3)+p_3(A_1,F_2,A_3)+p_3(A_1,A_2,F_3)+$$
$$+...$$

$$=\sum _{k=1}^{\infty }(\sum _{j=1}^{k}p_k(A_1,...,A_{j-1},F_j,A_{j+1},...,A_k))$$
  We shall now rewrite Polyakov's
formula using iterated integrals and we shall see that it coincides
with the above expression.

 The key point in Polyakov's formula is that he actually "breaks" the loop at a point $s$ and
then integrates over all $s$, namely from 0 to 1. We know how to write
the path order exponent using iterated integrals. However one now must
distinguish between two simplices because we have broken the loop. 

In all of our discussion above, the integrals were taken over the
standard simlpices over the interval [0,1], namely $\Delta _1$ is the
interval [0,1] and then $\Delta _k$ was $\{(s_1,...,s_k)\in {\bf R}^k | 0\leq
s_1\leq ...\leq s_k\leq 1\}$. We shall continue to keep this notation for the
simplices over the interval [0,1]. The same holds for the maps $p_k$.

Now we brake the loop at the point $s$, so we must, in addition, have two extra  classes of simplices:\\

I. One will be denoted $\Delta _k^s$ and the $\Delta _1^s$ will simply
be the interval [0,s] and the general $\Delta _k^s$ will be
$$\{(s_1,...,s_k)\in {\bf R}^k | 0\leq s_1\leq ...\leq s_k\leq s\}$$
The corresponding maps $p_k$ will be accordingly denoted $p_k^s$.\\

II. The other will be denoted $\Delta _k^1$ and the $\Delta _1^1$ will
simply be the interval [s,1]  and the
general $\Delta _k^1$ will be 
$$\{(s_1,...,s_k)\in {\bf R}^k | s\leq s_1\leq ...\leq s_k\leq 1\}$$
The maps will be denoted
$p_k^1$ in this case.

 Now with the above notation, Polyakov's variation can be written as:

$$\delta h=\int _0^1 ds (\sum _{n=0}^{\infty
}p_n^s(A_1,...,A_n))iF(s)(\sum _{n=0}^{\infty }p_n^1(A_1,...,A_n)$$ 
where we supress the indices and remember that the integral $ds$ refers
to the $F$ factor, indicated as $F(s)$.

 If we expand the above, we get:

$$\delta h=\int _0^1 ds
(1+p_1^s(A_1)+p_2^s(A_1,A_2)+...)iF(s)\times $$

$$\times (1+p_1^1(A_1)+p_2^1(A_1,A_2)+...)=$$

$$=\int _0^1ds
(iF(s)+iF(s)p_1^1(A_1)+iF(s)p_2^1(A_1,A_2)+...$$

$$+p_1^s(A_1)iF(s)+p_1^s(A_1)iF(s)p_1^1(A_1)+p_1^s(A_1)iF(s)p_2^1(A_1,A_2)+...$$

$$+p_2^s(A_1,A_2)iF(s)+p_2^s(A_1,A_2)iF(s)p_1^1(A_1)+p_2^s(A_1,A_2)iF(s)p_2^1(A_1,A_2)$$

$$+...)$$

 Now here comes algebraic topology to say that, in our notation

$$\int _0^1 ds\Delta _i^s*\Delta _j^1=\Delta _{i+j+1}$$
where the extra vertex $s$ goes in the $(i+1)$-st slot.

With the above in mind, the formula gives exactly  our expression for
$dh$, namely: 

$$\delta h=p_1(F_1)+p_2(F_1,A_2)+p_3(F_1,A_2,A_3)+...$$

$$+p_2(A_1,F_2)+p_3(A_1,F_2,A_3)+p_4(A_1,F_2,A_3,A_4)+...$$

$$+p_3(A_1,A_2,F_3)+p_4(A_1,A_2,F_3,A_4)+p_5(A_1,A_2,F_3,A_4,A_5)+...$$

$$+...$$

$$=\sum _{k=1}^{\infty }(\sum _{j=1}^{k}p_k(A_1,...,A_{j-1},F_j,A_{j+1},...,A_k))=dh$$

QED\\

\section{Remarks:}

 1. The appearence of the curvature $F$
is natural: the reason is Theorem 2 above: the
curvature has two terms, the $dA$ term comes from the $d$ part and the
$A\wedge A$ part comes from the $b_0$ part of the Hochschild
differential $b=d+b_0$. And it is the Hochschild cohomology of the
cyclic bar complex which is isomorphic with the de Rham cohomology of
the loop space.\\

 2. With our formalism, the flat connection on the loop space (cf
\cite{polyakov}, \cite{tsou}) is just

$$\tilde{A}=h^{-1}dh$$
where $h^{-1}$ is expressed exactly like $h$ using sum of iterated
integrals but now the vertices of the simplices will be in the
opposite order, namely 

$$0\leq s_n\leq ...\leq s_1\leq 1$$

One can see that this is exactly the standard expression for flat
connections for finite dimensional bundles. The proof that the above defined connection is flat now becomes
trivial, exactly like the finite dimensional case.\\

 3. Similarly the analogue of Yang-Mills equations for loop space
simplifies drastically.\\

 4. Finally, in virtue of Theorem 5 quoted above all the expresions are
reparametrisation invariant since we use iterated integrals.

\end{document}